\documentclass[twocolumn,aps,prb,graphicx,showpacs]{revtex4}

\usepackage{amsmath}
\usepackage{amsfonts}
\usepackage{amssymb}
\usepackage{graphicx}

\begin{document}

\title{Dynamical Mean Field Theory for transition temperature and optics of
pseudocubic manganites.}

\author{B. Michaelis$^{1}$ and A. J. Millis$^{2}$}

\affiliation{$(1)$ Department of Physics and Astronomy, Rutgers University\\
136 Frelinghuysen Rd, Piscataway NJ 08854\\
$(2)$ Department of Physics, Columbia University\\
538 W. 120 St, NY NY 10027} 
\date{\today}

\begin{abstract}

A tight binding parametrization of  local spin density functional band
theory is combined with a dynamical mean field treatment of correlations to
obtain a theory of the magnetic transition temperature  and optical
conductivity and $T=0$ spinwave stiffness
of a minimal model for the pseudocubic metallic $CMR$
manganites such a $La_{1-X}Sr_{x}MnO_{3}$. The results indicate that
previous estimates of $T_{c}$ obtained by one of us
(Phys. Rev. \textbf{B61} 10738-49 (2000)) are in error,
that in fact the materials are characterized by Hund's coupling
$J\approx 1.5eV$, and that magnetic-order driven changes
in the kinetic energy may not be the cause of the observed
'colossal' magnetoresistive and multiphase behavior in the manganites,
raising questions about our present understanding
of these materials.

\end{abstract}

\maketitle


\section{Introduction}
Pseudocubic manganites such as $La_{1-x}Sr_{x}MnO_{3}$ are of continuing
interest to physicists and materials scientists for many reasons, including
the possibility that the materials are 'half metals', in other words, have
ferromagnetic ground states with fully spin polarized conduction bands. The
possibility of 'half metallicity' was understood in the 1950s by Wollan,
Kohler, Zener, Goodenough, and Anderson
\cite{Zener51,Wollan55,Anderson55,Goodenough55}, but remains 
controversial today, with some calculations \cite{Pickett96} and some \cite{Soulen}
(but not all \cite{Park98}) experiments suggesting the presence of minority carriers.

In the pseudocubic manganites such
as $La_{1-x}Sr_xMnO_3$ the possible half metallicity 
arises from correlation and crystal-field effects 
involving the $Mn$ d-levels, which in a naive formal-valence
picture contain $4-x$ d-electrons (see e.g Ref \cite{Wollan55}
for further discussion).
In brief, in these materials the nearly cubic crystal field splits the d-multiplet
into a lower-lying $t_{2g}$ symmetry triplet and a higher lying $e_g$-symmetry doublet.
Correlation effects cause the $t_{2g}$ levels to contain three electrons with
parallel spin, producing an electrically inert $S=3/2$ 'core spin', while the remaining
$1-x$ electrons go into $e_g$-symmetry orbitals and can move throughout the crystal
subject to a Hund's coupling $J$ which favors configurations in which an $e_{g}$
electron on a given site has its spin parallel to the core spin on that 
site \cite{oxygen} Evidently, if the Hund's 
coupling is strong enough, then a state in which all of the core 
spins are aligned will also have a fully polarized conduction band; 
therefore the crucial question becomes: what is the value of $J$? 

From the earliest days \cite{Zener51,Anderson55} through the revival of
interest in the mid 1990s \cite{Millis96b} and including recent literature
\cite{Evertz2002a,Evertz2003,Hotta01}, many workers have assumed that $J$ is much larger
than the electronic bandwidth. Such a large value of $J$ has
been argued to lead to interesting observable 
effects including a temperature dependent optical  spectral weight
in the physically
relevant $\omega<4eV$ regime
\cite{Millis96b}  and to provide a direct connection
between optical conductivity and the values of the magnetic transition
temperatures \cite{Chattopadhyay00}. A temperature
dependent optical spectral weight has indeed been observed \cite{Quijada98}.
Even more importantly, the experimental phenomenology \cite{Quijada98}
suggests that there
is a strong connection between electronic energetics and magnetic order
and that in particular changes in electron kinetic energy occurring as the system
is driven through ferromagnetic-non-ferromagnetic transitions lead to the
observed \cite{Cheong98} multiphase behavior, suggesting again a very large
value of $J$. On the other hand, Measurements
on gas-phase $Mn$ \cite{Sugano70} suggest a relatively small $J$ (comparable
to the electronic bandwidth), and measurements on related transition metal
oxides \cite{Hillebrecht84,Hillebrecht85} 
suggest that gas-phase values for $J$ are only weakly renormalized
by solid-state effects. On the other hand, many workers
(including most of the references listed above)
interpret experiments in the 'double-exchange' picture which requires
a $J$ much larger than the bandwidth.
Determining the magnitude of $J$ and the effect of 
magnetic order on energetics is
therefore an important issue, which remains controversial.

Tokura and co-workers noted \cite{Tokura97} that the optical
conductivity may contain  direct information about
the value of $J$: in  a paramagnetic state optical processes were possible in
which an electron  moves from one site, where it was locally parallel
to the core spin, to an  adjacent site in which (because the core
spin is differently oriented) it is  locally antiparallel. Such
transitions should occur at an energy set by the  Hund's coupling $J$,
and should become visible as the temperature is raised. 
Ref. \cite{Chattopadhyay00} analysed this issue in detail,
presenting a   comparison between theory and
experiment which was argued to exclude  the possibility that the
`antiparallel` transitions  had been observed.

In this paper 
we reexamine the previous work \cite{Millis96b,Chattopadhyay00}, 
using a more realistic tight-
binding parametrization of the relevant portions of the band structure
in combination with  a dynamical mean field treatment of the interaction 
between electrons and  core spins and employing more detailed
and extensive numerics.  While various aspects of this physics
have been extensively studied, especially for model systems, 
we believe that this paper is the first to combine
a realistic band parametrization with a complete treatment
of {\it dynamical} and thermodynamic quantities
such as the conductivity (for both low-T and $T>T_c$), spin-wave stiffness
and transition temperature.
(Of course,  many important 
studies of {\it static} properties using realistic bands have appeared: 
for a recent example
see \cite{Popovic02}.  We also
note that in early important work Takahashi and Shiba
computed $T=0$ optical conductivities using a similar 
parametrization of band theory \cite{Takahashi98}.)
We compute magnetic transition 
temperatures, optical  conductivities and optical spectral weights
and the $T=0$ spin-wave stiffness for a range of model parameters  
and by comparison to experiment 
estimate the actual physical parameters for  the material.    
We find that contrary to statements made in some
previous work \cite{Chattopadhyay00} a moderate value of $J$ provides
a reasonable account of many aspects of the optical spectrum
and a plausible account of the magnetic transition temperature
and $T=0$ spin wave stiffness.
The differences with previous work apparently arise
because  the calculations  of Ref
\cite{Millis96b,Chattopadhyay00} were performed on 
perhaps oversimplified model systems. Also some of the results
\cite{Chattopadhyay00} contain numerical errors.

\section{Model}    
\subsection{\protect\bigskip Hamiltonian}    
We consider electrons moving among sites of a simple cubic lattice   
according to a tight binding model with hopping matrix elements obtained  
from band theory and coupled to a local classical 'core spin'.  Thus we  write:   
 
\begin{equation} H=H_{band}+H_{int}  \label{H} 
\end{equation}    

For the interaction we take a local coupling of core spins to  conduction electrons:   
\begin{equation} 
H_{int}=-J\sum_{a,b,\alpha, \beta ,i}\overrightarrow{S}_{i}\cdot c_{a,i,\alpha }^{+} 
\overrightarrow{\sigma }_{\alpha \beta }c_{a,i,\beta }.  
\label{hint} 
\end{equation}    

We obtain $H_{band}$ from LSDA band calculations \cite{Pickett96,Satpathy97}, 
which are  
consistent with straightforward quantum chemical (Goodenough-Kanamori)  
considerations. Both approaches suggest that the itinerant electrons lie 
in  two symmetry related bands which may be thought of as arising from  
electrons hopping among $e_{g}$ symmetry local orbitals and 
are  reasonably well described by a nearest neighbor tight binding model 
\cite{Ahn00,Chattopadhyay00} . To write this hopping it is convenient 
to adopt 
a  Pauli matrix notation in which the up state is the $|x^{2}-y^{2}>$ orbital 
and the down state is the $|3z^{2}-r^{2}>$ orbital. The basic hopping is  
orbital dependent so that   

\begin{equation} H_{band}=\frac{-1}{2}\sum_{i\delta ab\alpha }
(t(\delta )^{ab}c_{i+\delta,a,\alpha }^{+}c_{ib\alpha }+H.c.) 
\label{hband} 
\end{equation}  

where $\delta $ is a vector connecting site $i$ to a nearest neighbor 
and  the hopping is determined by a 2x2 matrix given after Fourier 
transformation  by    

\begin{equation} 
\mathbf{\varepsilon }=\varepsilon _{0}(p)+
\overrightarrow{\varepsilon }(p)\cdot  \overrightarrow{\tau }  
\label{e} 
\end{equation} 

with $\overrightarrow{\tau }$ the usual Pauli matrices, $b$ the lattice constant,
  
\begin{equation} \varepsilon _{0}(p)=-t(\cos (p_xb)+\cos (p_yb)+\cos (p_zb))  
\label{e0} 
\end{equation} 

and $\overrightarrow{\varepsilon }(p)=(\varepsilon _{x}(p),0,\varepsilon _{z}(p))$ with 

\begin{eqnarray} \varepsilon _{x}(p) &=&-\frac{\sqrt{3}t}{2}(\cos (p_xb)-\cos (p_yb))  
\label{ex} \\ 
\varepsilon _{z}(p) &=&t(\cos (p_zb)-\frac{1}{2}(\cos (p_xb)+\cos (p_yb)) 
\label{ez} 
\end{eqnarray}    

The energy eigenvalues are 

\begin{equation} E_{\pm }=\varepsilon _{0}\pm \sqrt{\varepsilon _{x}^{2}+
\varepsilon _{z}^{2}} 
\label{eplusminus} 
\end{equation}   

Note that along the zone diagonals ((1,1,1) etc)
$\varepsilon _{z,x}=0$ so  the two bands are degenerate and along the
line to any cube face ((1,0,0)  and equivalent) one of the two bands
is dispersionless. These two features  occur, to a high degree of
accuracy, in the calculated band structure,  lending support to the
nearest neighbor tight binding modelling. 

The best fit to the calculated band theory leads to $t=0.67eV$. This, 
for  example, implies that at band filling $n=0.7$ the density of  
states per $Mn$-ion is $0.57/eV$ which reproduces   almost exactly 
the local spin density approximation
value $0.58/eV$ per $Mn$-ion quoted on
p. 1154 of Ref \cite{Pickett96}. (Note also that \cite{Pickett96} 
contains a misprint  \cite{Pickett00} in  the value of the Drude plasma 
frequency $\Omega_{p}=1.9eV$. The correct band theory value is  very 
close to the $\Omega_p=\sqrt{4 \pi e^2 D/b}=3.1eV$  which follows using 
$t=0.67eV$ in Eq. \ref{Ddef} below). 

\begin{figure}[t]
\includegraphics[angle=-90,width=3.5in]{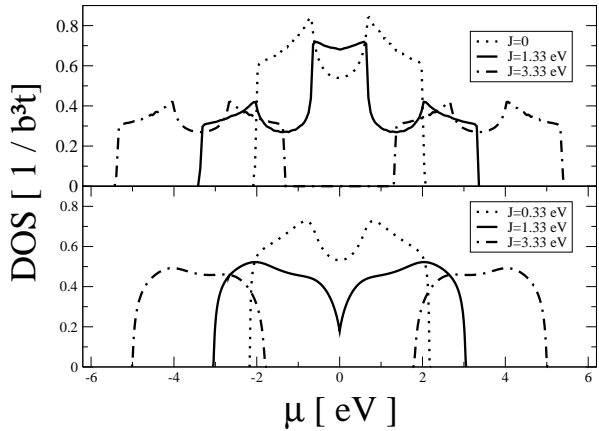}

\caption{{\it Upper panel}: $T=0$ density of states for a ferromagnetic
ground state, plotted against chemical potential for several different
Hund's couplings $J$.
{\it Lower panel}: Density of states in 
paramagnetic phase plotted against 
energy for several $J$} 
\end{figure}  

The energy bands extend from $-3t$ to 
$+3t$. Useful quantities to  characterize the state of the system 
include the particle density  
  
\begin{equation} 
n=\sum_{a,\sigma }<c_{ia\sigma }^{+}c_{ia\sigma }>  
\label{nofmu} 
\end{equation}   

the density of states 
\begin{equation} D(\epsilon)=\frac{\partial n}{\partial \mu} 
\label{dos} 
\end{equation}
(whose $T=0$ value is shown for different $J$ values in the 
upper  panel of Fig. 1), the 'kinetic energy' per site $K=-\frac{<H_{band}>}{N}$ ($N$ is the number
of sites) given by
\begin{equation} 
K=\frac{1}{2N}\sum_{\mathbf{\delta }a\sigma i}t^{ab}(\mathbf{\delta })
<c_{ia\sigma }^{+}c_{i+\mathbf{\delta }a\sigma }+H.c.>  
\label{Kdef} 
\end{equation}  
and the 'Drude weight' $D$ which for a spin polarized system and current
flowing in the $z$ direction is
\begin{equation} 
D_{z}=b^{3}\int \frac{d^{3}p}
{\left( 2\pi \right) ^{3}}\sum_{\lambda =+,-}
\left( \frac{\partial E_{\lambda }(p)}{b^2\partial p_{z}}\right) 
^{2}\delta (E_{\lambda }(p)-\mu )  
\label{Ddef} 
\end{equation}      

Figure 2 shows the kinetic energy (divided by 3 to  facilitate comparison to 
optics) and Drude weight,  plotted against particle density for  a fully 
spin polarized gound state.    

\begin{figure}[b]  
\includegraphics[angle=-90,width=3.5in]{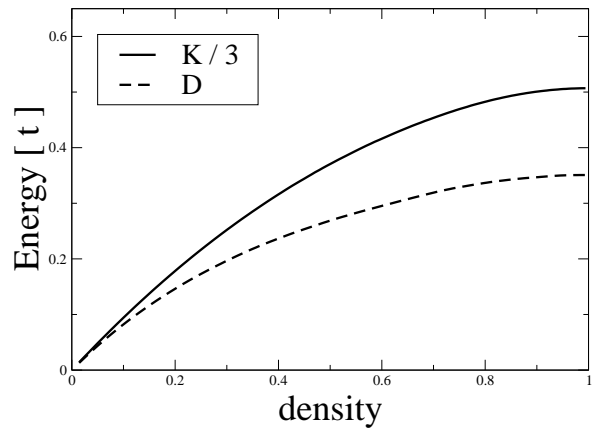}
\caption{Solid line: Kinetic energy (from Eq \ref{Kdef} but
divided by 3) in units of hopping parameter $t$ plotted
vs. particle density $n$.  Dashed line: Drude weight $D_z/t$ (from Eq \ref{Ddef})}
\end{figure}      

\subsection{Conductivity}    To obtain the conductivity we require a coupling 
between the electric field  and the electronic states. We represent the 
electric field by a vector  potential $A$ and adopt the Peierls phase approximation, 
$t_{i-j}\rightarrow t_{i-j}e^{i\frac{e}{c}\overrightarrow{A}\cdot \overrightarrow{R}_{ij}}$.  
This approximation has been argued to be accurate  in other transition metal oxide 
contexts \cite{Ahn00,Millis00} and was also used
by Takahashi and Shiba \cite{Takahashi98}.
The  current \ density operator in the $z$ direction is    

\begin{equation} \widehat{J}_{z}\equiv \frac{\delta \widehat{H}}
{Nb^3\delta A_{z}}=-\frac{2te}{b^{2}}\left( \sin (p_{z}b)-\frac{eb}{c}A\cos (p_{z}b)\right) 
\left(  
\begin{array}{cc} 0 & 0 \\  0 & 1 
\end{array} \right)  
\label{jz} 
\end{equation}  

The expectation value of the term in $J$ proportional to $A$ 
gives the total  
oscillator strength, $S(\infty )$ in the 
conduction band contribution to the  
optical conductivity (see \cite{Maldague76,Quijada98,Millis00} for details).  
We have 
  
\begin{equation} 
S(\infty )=\frac{e^{2}}{3b}K  
\label{S} 
\end{equation}    

The conductivity is  
\begin{equation} 
\sigma (\Omega )=\frac{S(\infty )-\chi _{jj}(\Omega )}{i\Omega }  
\label{sig} 
\end{equation}  
with $\chi _{jj}$ the usual Kubo formula current-current correlation  
function evaluated using $J$ evaluated at $A=0$.   

\subsection{Spin Wave stiffness}
To compute the $T=0$ spin wave stiffness we follow the standard procedure
outlined in \cite{Millis95,Millis99} We compute the energy cost of a small 
amplitude, long wavelength
rotation of the order parameter away from the fully polarized ferromagnetic
state, which we take to be 
aligned with the $z$ axis. The calculation is
most easily carried out by locally rotating the spin 
quantization axis to
align with the local spin direction, so 
that the $T=0$ Hamiltonian becomes
(note repeated orbital and spin indices $a,b,\alpha,\sigma,\sigma'$
are summed over)

\begin{equation}
H=-\frac{1}{2}\sum_{i,\delta }t^{ab}(\delta )
c_{i+\delta ,a,\sigma }^{+}c_{i,b\sigma^{\prime }}
R_{i+\delta \sigma \alpha }^{+}R_{i\alpha \sigma ^{\prime }}+H.c.
\label{H-sw}
\end{equation}
where $R$ are the usual $S=1/2$ rotation matrices. We find two terms.  The
first one arises from the term in $R^{+}R$ proportional to the square of the
deviation, $\overrightarrow{m}_{q}$\ of the magnetization from its ordered
state value and is ($K$ is the kinetic energy defined above in Eq \ref{Kdef}).

\begin{equation}
E^{(1)}=\frac{K}{24}\left( qa\right) ^{2}\overrightarrow{m}_{q}\cdot\overrightarrow{m}_{-q}  
\label{E1}
\end{equation}

while the second one arises from inserting the expression for the 
term in $R^{+}R$ which is linear in $\overrightarrow{m}_{q}$ 
into the familiar second order perturbation theory expression, and is
\begin{equation}
E^{(2)}=-\left( qa\right) ^{2}\overrightarrow{m}_{q}\cdot \overrightarrow{m}
_{-q}I_{SW}(J) 
\label{E2}
\end{equation}
with
\begin{equation}
I_{SW}(J) =\sum_{p}t^{2}\sin ^{2}(p_{z})\left(\Psi_+(p)+\Psi_-(p) \right)
\label{ISW}
\end{equation}
where (note we have suppressed the momentum labels to avoid 
clutter in the equations)
\begin{equation}
\Psi_+= \frac{f(E_{-})\left( J\cos ^{2}(\theta )+\cos ^{4}(\theta )
\sqrt{\varepsilon _{x}^{2}+\varepsilon _{z}^{2}}\right) }
{2J\left( J+\sqrt{\varepsilon _{x}^{2}+\varepsilon _{z}^{2}}\right) }
\label{Psiplus}
\end{equation}
\begin{equation} 
\Psi_-=\frac{f(E_{+})\left( J\sin ^{2}(\theta )-\sin ^{4}(\theta _{p})
\sqrt{\varepsilon
_{x}^{2}+\varepsilon _{z}^{2}}\right) }
{2J\left( J-\sqrt{\varepsilon
_{x}^{2}+\varepsilon _{z}^{2}}\right) }  
\label{Psiminus}
\end{equation}
and 
$\tan (\theta )=\varepsilon _{x}/\left( \sqrt{\varepsilon
_{x}^{2}+
\varepsilon _{z}^{2}}+\varepsilon _{z}\right) .$ 

The spin wave stiffness $D_{SW}$ as conventionally defined is given by twice the
total coefficient of $\left( q\right) ^{2}\overrightarrow{m}_{q}\cdot 
\overrightarrow{m}_{-q}$ divided by the ordered moment, i.e.
\begin{equation}
D_{SW}=\left[\frac{K}{12}-2I_{SW}(J)\right]b^2 
\label{D}
\end{equation}
This expression  applies to the tight
binding band structure and to all $J$ such that the 
ground state is fully polarized. It is consistent  with those obtained by 
Kubo and Ohata \cite{Kubo72} 
and Furukawa \cite{Furukawa96}
who considered simpler models in the infinite $J$ and
(Furukawa) first $1/J$ correction.
We have evaluated $D$ (Eq \ref{D}) for the chemical potential
$\mu=-1$ corresponding to the widely studied $0.7$ doping;
results, made dimensionless by dividing by $tb^2$ are shown in Fig. 3.
\begin{figure} 
\includegraphics[width=3.0in]{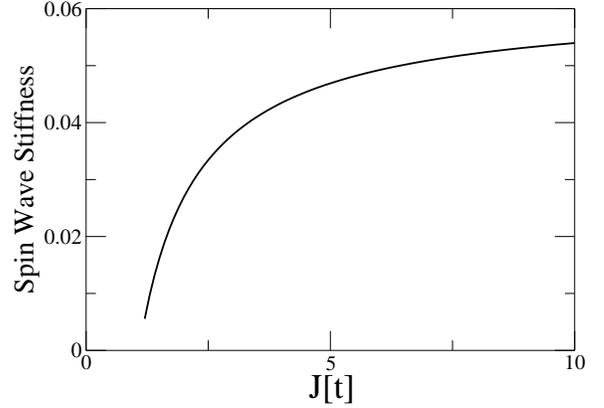}  
\caption[3]{Dimensionless spin wave stiffness $D_{SW}/(tb^2)$
as function of Hund's coupling $J$
divided by bandwidth $t$, for $T=0$ fully polarized state of
two-orbital lattice model at carrier concentration
$n=0.7$, calculated using Eq. \ref{D} } 
\end{figure}

\section{Method of Evaluation}  
\subsection{Overview}  
To evaluate the non-ground-state
properties of $H$ we use the dynamical mean field 
method\cite{MetznerVoll89,muellhartmann89a,Georges96,Chattopadhyay00} This method 
is extensively described and  justified elsewhere, and is relevant 
here because the principal interactions  are local. In brief the 
central approximation concerns the electron Green  function. For the 
band structure of present relevance the electron Green  function 
is written in general as    

\begin{equation} 
\mathbf{G(}z,p)=(z-\mathbf{\Sigma }(p,z)-{\bf \varepsilon }(p)+\mu )^{-1} 
\label{gdef}
\end{equation}  

with $\Sigma (p,z)$ the self energy. In the dynamical mean field method one  
takes $\Sigma $ to be $p-$independent, i.e. to depend only on $z$. 
The  important quantity is then the momentum integrated Green function    

\begin{eqnarray} 
G_{mom-int}(z) &=&b^{3}\int \frac{d^{3}p}{\left( 2\pi \right) ^{3}} {\bf G(z,p) }
\label{gmomint} 
\end{eqnarray}  

The local Green  function, being a function of frequency only, may be 
derived from a local  quantum field theory, which is specified by a 
partition function $Z_{loc}$  given in terms of a mean field function 
$a(\tau )$. This is in general a  matrix in orbital and spin indices 
(which we do not explicitly write here)  
  
\begin{eqnarray} 
Z_{loc}&=&\int \mathcal{D}c^{+}cExp[\int d\tau d\tau ^{\prime }c^{+}(\tau )
{\bf a}(\tau -\tau ^{\prime })c(\tau ^{\prime }) \nonumber \\
&+&\int d\tau H_{int}]  
\label{zloc} 
\end{eqnarray}    

>From this action one may extract a local Green function $G_{loc}$ 
and self  energy $\Sigma $ via   
 
\begin{equation} 
G_{loc}(\tau )=\frac{\delta ln Z_{loc}}{\delta a(\tau )}
=\left( a-\Sigma \right) ^{-1}  
\label{gloc} 
\end{equation}    

The mean field function $a$\ is fixed via the constraint that the local  
Green function calculated from $Z_{loc}$ is identical to the momentum  
integrated Green function obtained from Eq \ref{gmomint}, using the local  
self energy defined in the second equality in Eq \ref{gloc}, i.e.  
  
\begin{equation} 
G_{loc}(z )=G_{mom-int}(z)  
\label{sce} 
\end{equation}    

In the present problem the two orbitals are degenerate so it is not  
necessary to consider orbital indices in $G_{loc}$.    For T=0K a 
ferromagnetic core spin configuration yields a spin dependent but  
frequency independent $\Sigma _{\sigma }=\sigma _{z}J$. 
For $T>T_{c}$ Eq   \ref{sce} is an integral equation for 
$\Sigma $, which is solved by  numerical iteration.    

\subsection{\protect\bigskip Conductivity} 
In the dynamical mean field approximation there are no vertex 
corrections for the  current current correlation function \cite{khurana,Georges96}, 
so it is given by 

\begin{eqnarray} 
\chi _{jj}(i\Omega )&=&T\sum_{\omega _{n}}\int \frac{d^{3}p}{\left( 2\pi \right) ^{3}}
\nonumber \\
&&Tr\left[ \mathbf{J_{z}}\mathbf{G}(p,i\Omega +i\omega _{n}) 
\mathbf{J_{z}}\mathbf{G}(p,i\omega _{n})\right]   
\label{chijj} 
\end{eqnarray}  
with $G$ given by Eq. \ref{gdef} using the momentum independent self energy
from the final solution of Eq \ref{gloc}.

The Matsubara sum and analytical continuation may easily be performed, and 
the dissipative part of the conductivity is given by 
$\sigma (\Omega )= Im[\chi _{jj}(\Omega -i\delta )]/\Omega $. 
It is convenient to transform within the trace to a orbital basis 
which diagonalizes the Green functions. In the model considered here 
the self energy is the same for both orbitals, and so we may split 
the result into an intra- and interband part, writing 
   
\begin{eqnarray} 
\sigma _{inter/intra}(\Omega )&=&
\int \frac{d\omega }{\pi }\theta(\omega,\Omega) \int 
\frac{d^{3}p}{(2\pi )^{3}} \nonumber \\ 
&&\frac{t^{2}sin^{2}(p_{z})\varepsilon _{x}^{2}}
{\epsilon _{x}^{2}+\epsilon _{z}^{2}} \Phi_{inter/intra}   
\end{eqnarray}  
with     

\begin{figure} 
\includegraphics[width=3.0in]{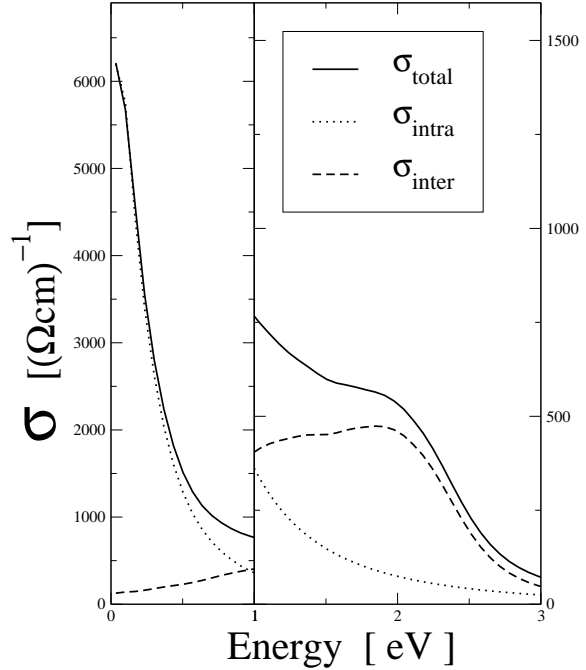}  
\caption[4]{Optical conductivity for the ferromagnetic ground state at n=0.7, 
computed as described in the text using $t=0.67eV$ and
a single-particle scattering rate $1/ \tau=0.1eV$. 
Note the change in $y$-axis scale at 
$1eV$.} 
\end{figure}

\begin{eqnarray} 
\theta(\omega,\Omega)&=&\frac{f(\omega)-f(\omega +\Omega )}{\Omega } \label{theta} \\ 
\Phi _{intra} &=&\frac{\varepsilon _{x}^{2}A^{(1)}(\omega )A^{(1)}(\omega+\Omega )}
{(\epsilon _{z}+\sqrt{\epsilon _{x}^{2}+\epsilon _{z}^{2}})^{2}} \nonumber \\ 
&&+\frac{\varepsilon _{x}^{2}A^{(2)}(\omega )A^{(2)}(\omega +\Omega )}
{(\epsilon _{z}-\sqrt{\epsilon _{x}^{2}+\epsilon _{z}^{2}})^{2}} 
\label{Phiintra} \\ \Phi _{inter} &=&A^{(1)}(\omega )A^{(2)}(\omega +\Omega ) \nonumber \\ 
&&+A^{(2)}(\omega )A^{(1)}(\omega +\Omega )   
\label{Phiinter} 
\end{eqnarray} 
The quantities $A$, $\varepsilon$ and $\Phi$ depend on momentum $p$ which we
have not written explicitly to avoid confusing clutter in the equations:
for example, the band-n ($n=1,2$) spectral function 
$A^{(n)}(p,\omega )=Im[\Sigma (\omega )]((\omega
-E_{n}(p)-Re[\Sigma (\omega )])^{2}+Im[\Sigma (\omega )]^{2})^{-1}$.

At $T>T_{c}$ the conductivity is calculated using the self energy
obtained from the solution of the mean field equations. 
For $T\rightarrow 0$ the spin-induced self energy vanishes (except for the
spin-dependent energy shift) and we have added a modest impurity broadening 
(scattering rate $ \Sigma''(\omega)=\tau ^{-1}=0.1t$) so the 
Drude peak is visible on the same
scale as the interband term. The calculated $T=0$ conductivity is shown in Fig. 4.

\subsection{Transition Temperature} To determine the transition 
temperature it is convenient to write the mean field function in the 
presence of a uniform magnetization \textbf{m}, which is in principle 
a matrix in spin space, as 

\begin{equation} 
\mathbf{a}=a_{0}\mathbf{1}+a_{1}\mathbf{\sigma }\cdot\mathbf{m} 
\label{aansatz} 
\end{equation} 

and to take the
magnetization direction to be parallel to $z$ so that after
integrating out the fermions we obtain for the impurity model action
(the two is for orbital degeneracy)    

\begin{equation}
S_{imp}=2\sum_{n}Tr\ln [a_{0}+a_{1}\sigma
_{z}+J\overrightarrow{S}\cdot  \overrightarrow{\sigma }]
\label{simp1} 
\end{equation} 

We can evaluate the trace over spins,
getting ($\theta $ is the angle between the core spin direction and
the magnetization direction, which is taken to be the $z$ direction)

\begin{equation} 
S_{imp}(\theta )=2\sum_{n}
\ln[a_{0}^{2}-a_{1}^{2}-J^{2}S^{2}-2a_{1}JS\cos (\theta )]
\label{simp2} 
\end{equation} 

Assuming (as detailed studies of
simpler model systems have confirmed \cite{Millis96b}) that the transition is second
order we may study it by linearizing in $a_{1}$, so that the
magnetization $m$ is given by 

\begin{eqnarray} 
m &=&{\large <}\frac{\int d\cos \theta \left( \cos (\theta )
Exp[S(\theta)]\right) }{\int d\cos \theta Exp[S(\theta )]}{\large >}  \notag
\\ &=&-\frac{4JS}{3}\sum_{i\omega_{n}}\frac{a_{1}}{a_{0}^{2}-J^{2}S^{2}}
+O(a_{1}^{3})  
\label{tc1}
\end{eqnarray}
    
The impurity (local) Green function corresponding to Eq
\ref{simp2} is 
 
\begin{eqnarray} 
G_{loc} &=&\langle \left(
    a_{0}+a_{1}\sigma _{z}+J\overrightarrow{S}\cdot
    \overrightarrow{\sigma }\right) ^{-1}\rangle  \nonumber \label{gimp1} \\
  &=&\frac{(a_{0}-\sigma _{z}[a_{1}(1+\frac{2}{3}\frac{J^{2}S^{2}}{
      a_{0}^{2}-J^{2}S^{2}})+Jm])}{(a_{0}^{2}-J^{2}S^{2})}+
\mathcal{O}(a_{1}^{2}) \nonumber 
\label{gimp2}\\ 
\end{eqnarray}  
where the average has been taken over the action defined by Eq \ref{simp2}.    
From Eq \ref{gimp2} we may compute the self energy 
$\Sigma=-G_{loc}^{-1}+a_{0}+a_{1}\sigma _{z}$ finding 
$\Sigma =\Sigma _{0}+\Sigma_{1}\sigma _{z}$:  

\begin{eqnarray} 
\Sigma _{0} &=&\frac{J^{2}S^{2}}{a_{0}}  \label{sig1} \\ 
\Sigma _{1} &=&\frac{1}{3}\frac{J^{2}S^{2}}{a_{0}^{2}}a_{1}-
\frac{a_{0}^{2}-J^{2}S^{2}}{a_{0}^{2}}JSm  
\label{sig2} 
\end{eqnarray}  

Finally, $\mathbf{a}$ is fixed from  

\begin{eqnarray} 
G_{loc} &=&\frac{1}{2}Tr\int 
\frac{d^{d}p}{\left( 2\pi \right) ^{d}}
\left[\omega -\Sigma _{0}-\Sigma _{1}\sigma _{z}-
\mathbf{\varepsilon }_{p}\right]^{-1} \nonumber \\ 
&=&\frac{1}{2}Tr\int \frac{d^{d}p}{\left( 2\pi \right) ^{d}}
\left[ \omega-\Sigma _{0}-\mathbf{\varepsilon }_{p}\right] ^{-1}+ \nonumber \\ 
&&\left[ \omega -\Sigma _{0}-\mathbf{\varepsilon }_{p}\right] ^{-1}
\Sigma _{1}\sigma _{z}\left[ \omega -\Sigma _{0}-
\mathbf{\varepsilon }_{p}\right] ^{-1} \nonumber  \notag \\ 
&=&I_{1}(\omega -\Sigma _{0})+\Sigma _{1}\sigma _{z}I_{2}(\omega -\Sigma _{0}) 
\label{tc2} 
\end{eqnarray}  
with $I_{n}(z)=\int \frac{d^{d}p}{\left( 2\pi \right) ^{d}}
\left( \frac{1}{z-\varepsilon _{p}}\right) ^{n}$.

Eqs. \ref{tc1}-\ref{tc2} give an equation for T$_c$, which
contains only a$_0$ and the paramagnetic solution of Eq. \ref{sce}
which we denote here as $G_0$.
   
\begin{equation} 
T_{c}=-4J^2T_{c}\sum_{i\omega_n}
\left[\frac{G_0^2-I_2}{(3a_0^2-J^2)G_0^2+J^2I_2} \right]  
\label{tcfinal} 
\end{equation} 

This formula shows explicitly that in weak coupling
$T_c \propto J^2$.

\begin{figure}[t] 
\includegraphics[angle=-90,width=3.5in]{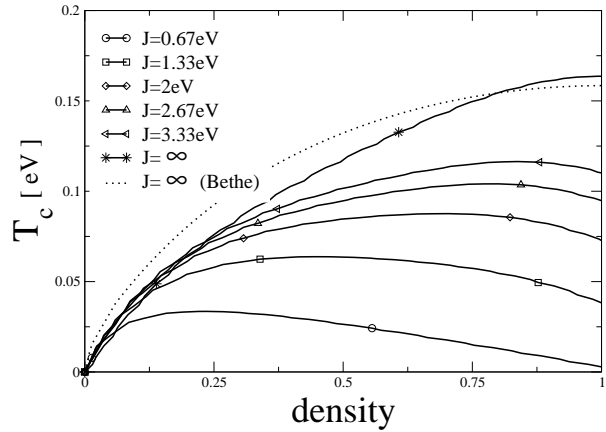} 
\caption{Calculated magnetic transition temperature plotted against 
band filling for several different $J$ values, 
along with $J=\infty$ Bethe-lattice result (for comparison to previous work).} 
\end{figure}

\section{Results}   

As noted above, the band
calculation is well fit by a $t$ of about $0.67eV$.  The physically
relevant densities correspond to $n<1$ and the most  experimentally
relevant density range is   $n=0.6-0.8$ where the ground state is in
fact a ferromagnetic metal. We have calculated the magnetic transition 
temperature as a  function of band filling for different values of the   
Hund's coupling $J$. 
Representative results are shown in   Fig. 5. We see that the calculated 
transition temperatures  become noticeably higher than the experimental 
range ($\sim$ 400K)  once $J$ becomes greater than a number of the order of  unity. Ref   
\cite{Chattopadhyay00} argued that the calculated transition temperatures for $J\gg t$  
were consistent with experimental data. 
We now believe that this conclusion was based on  
normalizing to an incorrect value for the kinetic energy and should 
be  disregarded. The present calculation reveals, in agreement with 
results  obtained by previous workers, that even modest values of 
$J$ suffice to push  transition temperatures into a range higher 
than the highest observed $T_{c}$. Corrections to 
the mean field approximation are believed \cite{Chattopadhyay00} in
this problem to reduce $T_{c}$ by of the order of $30\%$ and do not 
change this conclusion.

\begin{figure}[t]
\includegraphics[width=3.5in]{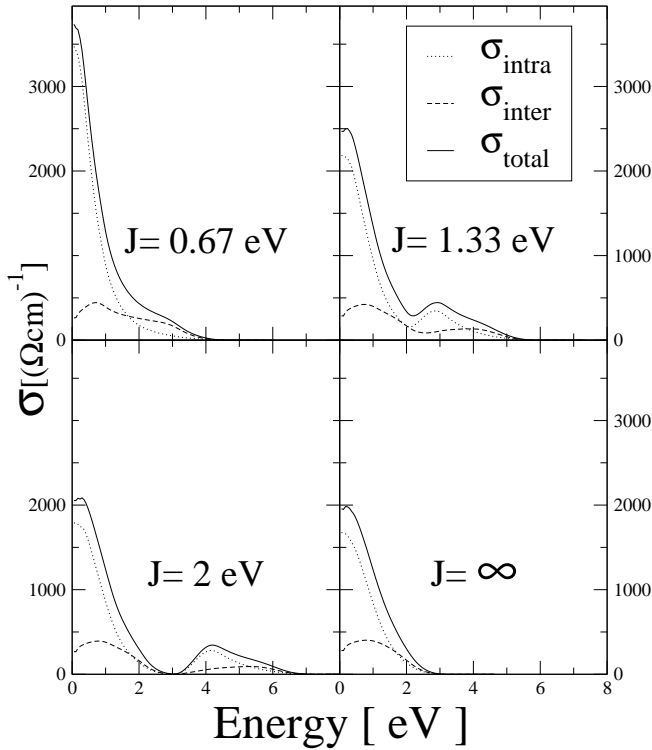}  
\caption[4]{Optical conductivity in the 
paramagnetic phase at ($T>T_c$) for different couplings $J$ and density
n=0.7, computed using $t=0.67eV$.} 
\end{figure}

The transition temperature data therefore suggest that a 
relatively modest $J$, corresponding to a local spin flip energy
$2J \sim 2-3eV$ and less than the full bandwidth $\sim 4eV$, might be appropriate
for the CMR manganite materials. Such a value of $J$ would be consistent with
the value $J \approx 2.7eV$ appropriate for gas-phase $Mn$ (note that spectroscopic
studies such as \cite{Hillebrecht84,Hillebrecht85} 
indicate that the Hund's coupling is much less renormalized
by solid-state effects than is the on-site Coulomb interaction). 
Our computed values for the $T=0$ spin wave stiffness also support
a lower value for $J$. Values in physical units may be obtained
by multiplying the data shown in Fig 3 by $tb^2 \approx 9.67$ 
(using the
numerical values $t =0.67 eV$ and $a=3.8 \AA$). A $J=2.5t$ yields a
$D=300meV\AA^2$, already much higher than the 
$D \approx 160-190eV\AA^2$ observed \cite{Baca98} 
in optimally metallic manganites.

Further evidence that
a smaller value of $J$ might be appropriate comes from our 
calculated optical conductivity,
shown in Fig. 6 for $T>T_c$ and various values of $J$.
Comparison to Fig 4 shows that if $J$ is larger than
about 1eV, then as $T$ is raised a new feature becomes visible at an 
energy of order $2J$. This feature arises  from motion of carriers 
from one site to another site with an antialigned  spin, and its existence 
was noted by Tokura and co-workers \cite{Tokura97}.  Such a feature 
is visible in experimental data at an energy of about $3eV$
\cite{Quijada98}. 
\begin{figure}[t]  
\includegraphics[width=3.0in]{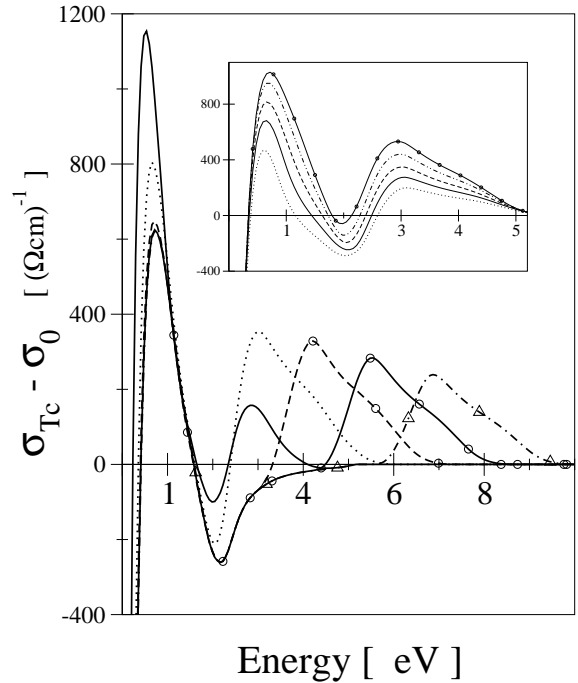}  
\caption{{\it Main panel:} difference 
$\sigma(\omega,T>T_c)-\sigma(\omega,T=0)$ for $n=0.7$
and $J=0.67eV$ (solid line) $1.33eV$ (dotted line) $J=2eV$ (long dashed line),
$J=2.67eV$ (second solid line) and $J=3.33eV$ (dot-dashed line).
{\it inset:} difference conductivity as function of density for
$J=1.33eV$ ($n=0.5$ (dotted), $n=0.6$ (solid) $n=0.7$ (dashed)
$n=0.8$ (dash-dot) and $n=0.9$ (second solid line). All curves computed
using $t=0.67eV$.}
\end{figure}  
Comparison of Fig. 7 to the data of \cite{Quijada98} shows 
that a   $J=2.5t \approx 1.5eV$ would  
approximately reproduce the observed peak
position and magnitude.  For smaller values of J 
the peak becomes difficult to distinguish
from the interband transition; for larger values the peak moves to higher
energies than is observed and the oscillator strength decreases below what
is observed. The Table shows the oscillator strength in the `` peak at $2J$'', 
obtained
by integrating $\sigma_{T_c}-\sigma_0$ from the highest zero-crossing up to 
$w=\infty$ and converting to kinetic energy via Eq 14 (e.g. the integration 
for $J=3.33eV$ and $n=0.7$ includes the area between 5.8eV and 9.6eV).  
It is also convenient to
define K$_{T_c}$,  the total spectral weight of $\sigma$ 
at T$_c$,  by evaluating Eqs \ref{Kdef}, \ref{S} at $T=T_c$. 
The ratio $K_{anti}/K_{T_c}$ can be compared with the experimental
results of Ref\cite{Quijada98}.
\begin{center} 
\begin{table}[b]
\caption{the kinetic energy ($\propto$ spectral weight), 
which corresponds to  the ``spin flip`` transition in the paramagnetic 
phase presented in absolute units and relative to the total $T=T_c$ spectral weight} 
\begin{tabular}{l||c|r} J[eV]  &  $-K_{anti}[meV]$  &  $K_{anti}/K_{Tc}$  
\\ \hline \hline 0.67&13.3&0.038\\ 1.33&55.6&0.17\\ 2.00&56.7&0.19
\\ 2.67&48.9&0.17\\ 3.33&42.2&0.15\\ \hline \hline 
\end{tabular} 
\end{table} 
\end{center}
Ref \cite{Chattopadhyay00} argued, on the basis of calculations based on
an oversimplified Bethe-lattice model,  that the observed oscillator strength
in the putative 'peak at $2J$' was too large to be consistent with the calculation.
The present calculations, which are based on the more realistic 
tight binding model, suggest that this
is not the case.

Previous work \cite{Millis96b,Chattopadhyay00} has drawn attention to a relationship
between optical spectal weight (i.e. kinetic energy) changes and the value of the magnetic
transition temperature. Our new results indicate that this result is of less generality
than suspected. Fig. 8 compares, for different values of $J$, the optical spectral weight
at $T=T_c$ (dashed and dotted lines) and $T=0$ (solid line). The 
$J=\infty$ limit reveals the expected approximately 30\% decrease in
spectral weight between $T=0$ and $T=T_c$, but the other traces reveal that the situation
is more complicated in general. Indeed, at $T=0$ and for $J$ sufficiently large, only one
spin band is occupied, and the band is therefore filled to a higher level, thereby
losing  kinetic energy. On the other hand,
at $T>T_c$ both spin directions are equally probable, so electrons can redistribute themselves
to the lower parts of the band, thereby gaining kinetic energy. 
However, in the paramagnetic
phase the non-vanishing
$J$ blocks some hopping processes. At intermediate $J$ these two processes compete, and
as seen in Fig. 8, which one is dominant depends on both the value of 
$J$ and the band-filling.
The blocking process is always dominant only at very large $J$.

\begin{figure}[t]  
\includegraphics[width=3.2in]{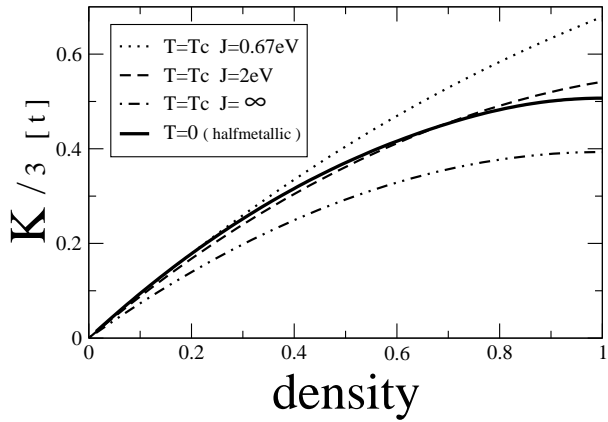}  
\caption{Comparison of optical spectral weight (expressed as a kinetic energy) 
at Tc (dashed and dotted lines) and 0 K (solid line) and various J; the 
minimal J value, which gives a halfmetallic ground state is dependent on n 
(e.g. for n=0.7 is J$_{min}$=0.64eV)}  
\end{figure}  

The non-systematic change of kinetic energy upon magnetic ordering observed for
reasonable values of $J$ suggests that the relation, proposed in previous work
\cite{Chattopadhyay00}, between change in kinetic energy and value of transition
temperature is not as general as expected. The lower panel of Fig. 9 shows the
value of the transition temperature, plotted against change in kinetic energy for several
different carrier densities, with $J$ as an implicit parameter.  Also shown, 
as the solid line, is the result obtained
in Ref \cite{Chattopadhyay00} (the different slope occurs because the
results of \cite{Chattopadhyay00} are for an orbitally non-degenerate model). 
A linear dependence seems 
to be reasonable and is consistent with the ideas 
in Ref.\cite{Chattopadhyay00} - but we observe,
that the slope is doping 
dependent. The upper panel in Fig. 9 shows the same plot for 
several different $J$, with carrier density as
an implicit parameter.  A much more complicated appearance is obvious.
We see that the relation between $T_c$ and kinetic energy is not at 
all universal, but depends strongly on 
electron density, essentially for the reasons given in the paragraph above.

\begin{figure}[t]  
\includegraphics[width=3.2in]{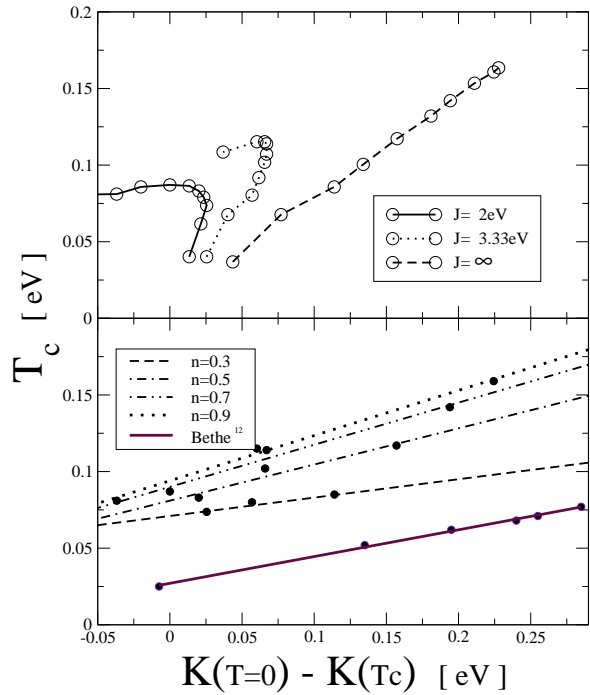}  
\caption{{\it Upper panel:} Open dots: calculated 
transition temperature plotted  against change in 
kinetic energy between $T=T_c$ and $T=0$ for different $J$
as shown with density $n$ as implicit parameter.
Dashed lines are guides to the eye. 
Along each line the doping increases by 0.1  
steps along the lines from 0.1 up to 1 or 0.9. 
{\it Lower panel:} Filled dots: calculated transition temperature
plotted against change in kinetic energy between $T=T_c$ and $T=0$
for different fixed densities, as shown, with Hund's coupling $J$
as implicit parameter. Dashed lines are results of a linear regression 
for the T$_c$ vs $\Delta K(J)$ 
curves. Solid line: results published in 
\cite{Chattopadhyay00} 
for a Bethe lattice; the difference seems to arise because 
of an omitted factor of two in
\cite{Chattopadhyay00}.}
\end{figure}   

\section{Conclusions}  We have presented a detailed analysis of
the optical conductivity and magnetic transition temperature of
electrons using a quasirealistic band structure and Hund's coupling to
classical 'core spins'. Our main finding is that, in contrast to
a widely-made assumption in the manganite literature but in agreement
with quantum chemical estimates and gas-phase measurements, a moderate
value of the Hund's coupling $J\approx 1.3-1.5eV=2.5t$
(large enough to fully polarize the 
conduction band at $T=0$ but 
less than the LSDA band width) suffices to account both for
the order of magnitude of the magnetic transition temperature and
for a feature observed in the optical conductivity in the paramagnetic
phase. We have also presented predictions for the evolution of this
feature as the carrier density is varied. Errors in previous work
\cite{Chattopadhyay00} (arising from use of an oversimplified band structure
and from numerical errors) have been corrected.
We have also presented results
for the spin-wave stiffness expected for a fully polarized ground state
and a tight-binding band structure at arbitrary $J$.

We observe that even with a modest $J$ the spin wave stiffness and
transition temperature are rather overestimated. It is possible to
fix up these discrepancies by adding a phenomenological antiferromagnetic
core-spin-core-spin coupling, as has been proposed by many authors.
However, in our view a more serious problem remains:
if these estimates for $J$ are accepted, then our basic understanding
of the physics of the manganites must be reconsidered, because for these
value of $J$ the change in kinetic energy between ferromagnetic and paramagnetic
state is very small. A series of experiments has made it clear
that the 'colossal' magnetoresistance is associated with a change in electronic
state from a more or less fermi-liquid like state to a state with
strong local lattice distortions, leading to either short ranged or
long ranged charge and orbital ordering. 
Previous work \cite{Millis96b}, based on a large-J limit, had predicted this
behavior as a consequence of the decrease in kinetic energy as spins were 
disordered, allowing the localizing tendency of the electron-phonon interaction
to overcome the banding tendency of the electrons and produce a new phase.
An approximately $40$ per cent decrease in 
conduction band oscillator strength, which is naturally associated to a decrease
in electronic kinetic energy, is indeed observed, and was argued to be consistent
with this physics. However, if J is in the range proposed here, then
the disordering of the spins does not significantly change the electronic kinetic
energy and another explanation for the high temperature second phase
and thus for the  'colossal' magnetoresistance must be sought. 

One possibility is that a 'Hubbard U' interaction is important, and reduces
the electronic kinetic energy and band width to the point where even the
modest $J$ we find here can have a strong effect. However, the total conduction
band spectral weight observed at low T is close to the full band theory
value, and indeed the while our simple calculation does not reproduce the
detailed lineshape, the calculated magnitude of the conductivity in the 1-3eV
is in reasonable agreement with data, suggesting that Hubbard-U effects are perhaps not
so important. It is possible, however, that the good agreement is accidental,
and that contributions from other orbitals (for example, 'wrong spin' $t_{2g}$
electrons) are important in the real materials, although they are absent
in our calculations.
We also note that the susceptibility of the finite J paramagnetic state to local 
(or long ranged) lattice distortions has not been computed. 
It might be that even though the kinetic energies of the
$T=0$ and $T>T_c$ phases are similar (at moderate J), the 
$T>T_c$ phase is more susceptible to local lattice distortions. This possibility
urgently deserves further investigation. 

\textit{Acknowledgements: }We thank G.A. Sawatzky, D. Singh 
and H.D. Drew for  helpful discussions and acknowledge  
support by the University of  Maryland-Rutgers MRSEC. B.M. additionally 
thanks the Stiftung der Deutschen  Wirtschaft (SDW) and the Wuerzburg-Rutgers exchange program.    


\begin{thebibliography}{99}
\bibitem{oxygen}  The actual situation is of course more complicated. 
In particular, what is commonly referred to as an $e_{g}$ electron 
is actually an $e_{g}$ symmetry antibonding combination of $Mn$ and $O$. 
While this distinction may for many purposes be neglected, it 
could be important for issues such as the value of interaction 
parameters.
\end{thebibliography}
\end{document}